\documentclass[sigconf]{acmart}
\usepackage{subcaption}
\usepackage{hyperref}
\usepackage{multirow, lipsum}
\usepackage{balance}

\usepackage{pifont}
\newcommand{\cmark}{\ding{51}}%
\newcommand{\xmark}{\ding{55}}%

\usepackage{hyphenat}

\newcommand{\orcidnew}[1]{\href{https://orcid.org/#1}{\includegraphics[width=8pt]{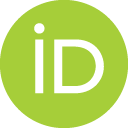}}}
\usepackage{xspace}
\usepackage{makecell}

\makeatletter
\DeclareRobustCommand\onedot{\futurelet\@let@token\@onedot}
\def\@onedot{\ifx\@let@token.\else.\null\fi\xspace}

\def\etal{\emph{et al}\onedot}

\AtBeginDocument{%
  \providecommand\BibTeX{{%
    \normalfont B\kern-0.5em{\scshape i\kern-0.25em b}\kern-0.8em\TeX}}}

\copyrightyear{2025}
\acmYear{2025}
\setcopyright{cc}
\setcctype{by}
\acmConference[GLSVLSI '25]{Great Lakes Symposium on VLSI 2025}{June 30-July 2, 2025}{New Orleans, LA, USA}
\acmBooktitle{Great Lakes Symposium on VLSI 2025 (GLSVLSI '25), June 30-July 2, 2025, New Orleans, LA, USA}
\acmDOI{10.1145/3716368.3735280}
\acmISBN{979-8-4007-1496-2/2025/06}

\begin{document}

\title{Emotion Detection in Older Adults Using Physiological Signals from Wearable Sensors}

\author{Md. Saif Hassan Onim \orcidnew{0000-0002-7228-2823}}
\affiliation{%
    \institution{University of Tennessee}
    \city{Knoxville}
    \state{TN}
    \country{USA}}
    \email{monim@vols.utk.edu}

\author{Andrew M. Kiselica \orcidnew{0000-0002-9514-0668}}
\affiliation{%
    \institution{University of Georgia}
    \city{Athens}
    \state{GA}
    \country{USA}}
    \email{akiselica@uga.edu}

\author{Himanshu Thapliyal \orcidnew{0000-0001-9157-4517}}
\affiliation{%
    \institution{University of Tennessee}
    \city{Knoxville}
    \state{TN}
    \country{USA}}
    \email{hthapliyal@utk.edu}

\renewcommand{\shortauthors}{Md. Saif Hassan Onim, Andrew M. Kiselica and Himanshu Thapliyal}

\begin{abstract}
Emotion detection in older adults is crucial for understanding their cognitive and emotional well-being, especially in hospital and assisted living environments. In this work, we investigate an edge-based, non-obtrusive approach to emotion identification that uses only physiological signals obtained via wearable sensors. Our dataset includes data from 40 older individuals. Emotional states were obtained using physiological signals from the Empatica E4 and Shimmer3 GSR+ wristband and facial expressions were recorded using camera-based emotion recognition with the iMotion's Facial Expression Analysis (FEA) module. The dataset also contains twelve emotion categories in terms of relative intensities. We aim to study how well emotion recognition can be accomplished using simply physiological sensor data, without the requirement for cameras or intrusive facial analysis. By leveraging classical machine learning models, we predict the intensity of emotional responses based on physiological signals. We achieved the highest 0.782 $r^2$ score with the lowest 0.0006 MSE on the regression task. This method has significant implications for individuals with Alzheimer’s Disease and Related Dementia (ADRD), as well as veterans coping with Post-Traumatic Stress Disorder (PTSD) or other cognitive impairments. Our results across multiple classical regression models validate the feasibility of this method, paving the way for privacy-preserving and efficient emotion recognition systems in real-world settings.
\end{abstract}

\ccsdesc[300]{Computing methodologies~Machine learning approaches}
\ccsdesc[500]{Human-centered computing~HCI design and evaluation methods}

\keywords{Emotion Detection; ADRD; Machine Learning; Wearables; TSST}
\maketitle


\section{Introduction} 

Understanding emotion is vital to understanding human behavior, especially in older adults, as emotional well-being is significantly related to cognitive and physical health. Research has shown that changes in emotional patterns can serve as early indicators of neurodegenerative diseases such as Alzheimer’s disease and related dementia (ADRD)~\cite{rhodus2024}. They also help diagnose mental health conditions such as PTSD, depression, and anxiety disorders in veterans and other vulnerable populations. Traditional emotion recognition approaches are primarily based on facial expression analysis, speech recognition, and behavioral assessments~\cite{din2024, badawi2024}. They often require camera-based monitoring or active user participation. Although effective, these methods raise concerns about privacy, intrusiveness, and real-world applicability. 

Recent advancements in wearable technology and physiological sensing have introduced a promising alternative to detect mental health conditions and relevant cognitive impairments~\cite{onim2023review, Delmastro2020, Kikhia2016}. Researchers employ both obtrusive and non-obtrusive instruments to gather physiological signals that serve as biometric indicators~\cite{onim2024utilizing, Cheong2020, Ferreira2014}. Beyond emotion recognition, previous studies have explored stress detection using these physiological signals. For example, research that uses EDA and HRV has shown promising results in identifying acute stress responses in real-time settings. Other studies have used Empatica E4 to assess the cognitive load, anxiety, and stress during educational tasks~\cite{onim2023casd, onim2024predicting, Belk2016}. These works reveal strong correlations between physiological changes and fluctuations in mental state and highlight the versatility of physiological sensors.

\begin{table*}[htbp]
\caption{Overview of Recent Literature on Emotion Detection from Physiological Signals}
    \centering
    \resizebox{2\columnwidth}{!}{
    \setlength{\tabcolsep}{10pt}
    \begin{tabular}{lccccc}
    \toprule
        \makecell[c]{\bf Authors \&\\ \bf Year} & \bf Data Acquisition & \makecell[c]{\bf Detected\\ \bf Emotions} & \makecell[c]{\bf Emotion\\ \bf Stimuli} & \makecell[c]{\bf Requires Video\\ \bf for Detection ?} & \makecell[c]{\bf Target Age Group\\ \bf of Participants}\\
                
    \midrule
        \makecell[c]{Wei~\etal~\cite{wei2017}\\2017} & EEG & \makecell[c]{Positive\\and Negative} & Pictures & \xmark & Adults\\
        
    \midrule
        \makecell[c]{Hui~\etal~\cite{hui2018}\\2018} & \makecell[c]{PPG, EDA,\\SKT and EMG} & \makecell[c]{Happiness, Anger, Fear,\\Disgust and Sadness} & Pictures and Videos & \cmark & Adults\\

    \midrule
        \makecell[c]{Quiroz~\etal~\cite{quiroz2018}\\2018} & HR & \makecell[c]{Neutral, Happy\\and Sad} & Musics and Videos & \xmark & Young Adults\\

    \midrule
        \makecell[c]{Heinisch~\etal~\cite{heinisch2018}\\2018} & \makecell[c]{EMG, EDA,\\TMP, EEG, BVP,\\RR and ECG} & \makecell[c]{HPVHA, HNVLA, HPVLA\\HNVHA and Neutral} & Pictures & \xmark & Young Adults\\

    \midrule
        \makecell[c]{Ragot~\etal~\cite{ragot2018}\\2018} & EDA and HR & Valence and Arousal & Pictures & \xmark & Young Adults\\
    
    \midrule
        \makecell[c]{Xiefeng~\etal~\cite{xiefeng2019}\\2019} & HS and ECG & \makecell[c]{Relaxed, Happy,\\Sad and Angry} & Musics and Videos & \xmark & Young Adults\\
        
    \midrule
        \makecell[c]{Costa~\etal~\cite{costa2019}\\2019} & \makecell[c]{EDA, PPG,\\IMU and GSR} & \makecell[c]{Anger, Contempt, Disgust,\\Fear, Happiness, Neutral,\\Sadness and Surprise} & Pictures & \cmark & Adults\\
        
    \midrule
        \makecell[c]{Jiang~\etal~\cite{jiang2019}\\2019} & EEG and Motion & \makecell[c]{Fear, Frustrated, Sad,\\Satisfied, Pleasant and Happy} & \makecell[c]{Musics, Pictures\\and Videos} & \cmark & Adults\\
        
    \midrule
        \makecell[c]{Zhang~\etal~\cite{zhang2020}\\2020} & \makecell[c]{HR, EDA, BVP,\\TEMP, PD,\\SA, and SV} & Valence and Arousal & Not Mentioned & \cmark & Young Adults\\

    \midrule
        \makecell[c]{Kim~\etal~\cite{kim2020}\\2020} & GSR and PPG & \makecell[c]{Pride, Elation, Joy,\\Satisfaction, Relief, Hope,\\Interest, Surprise, Sadness,\\Fear, Shame, Guilt, Envy,\\Disgust, Contempt and Anger} & \makecell[c]{Musics, Pictures\\and Videos} & \cmark & Adults\\
        
    \midrule
        \makecell[c]{Shu~\etal~\cite{shu2020}\\2020} & HR & \makecell[c]{Neutral, Happy\\and Sad} & \makecell[c]{Musics, Pictures\\and Videos} & \cmark & Young Adults\\
        
    \midrule
        \makecell[c]{Miranda~\etal~\cite{miranda2021}\\2021} & \makecell[c]{ECG, GSR,\\EEG and SKT} & Fear and Non-fear & Videos & \cmark & Adults\\
        
    \midrule
        \makecell[c]{Chen~\etal~\cite{chen2021}\\2021} & SP & \makecell[c]{Happiness, Sadness,\\Anger and Fear} & Videos & \xmark & Adults\\

    \midrule
        \makecell[c]{Badawi~\etal~\cite{badawi2024}\\2024} & \makecell[c]{PR, PRV, RR,\\ SCL} & \makecell[c]{Agitation\\and Aggression} & Indoor Activities & \cmark & Older Adults\\ 
    \midrule
        \makecell[c]{\bf Proposed Work\\ \bf 2025} & \makecell[c]{\bf EDA, BVP, IBI} & \makecell[c]{\bf Positive, Negative\\ \bf and Neutral} & \bf TSST Protocol & \xmark & \bf Older Adults\\
    \bottomrule

    \end{tabular}
    }
    \footnotesize{\\SA: Saccadic amplitude, SV: Saccadic velocity, HS: Heart Sound, PPG: Photoplethysmography, SKT: Skin Temperature, EMG: Electromyography, RR: Respiration Rate,\\PR: Pulse Rate, ECG: Electrocardiogram, EEG: Electroencephalography, EDA: Electrodermal Activity, SCL: Skin Conductance Level, IMU: Inertial Measurement Unit,\\PD: Pupil Dilation,HR: Heart Rate, TMP: Temperature, SC: Step Counter, BVP: Blood Volume Pressure, IBI: Inter Beat Interval, SP: Skin Potential, PRV: Pulse Rate Variability}
    \label{tab:comp}
\end{table*}

Table~\ref{tab:comp} provides a detailed overview of recent research on emotion detection using physiological signals, showcasing various methodologies, detected emotions, and participant demographics. Wei~\etal~\cite{wei2017} and Ragot~\etal~\cite{ragot2018} employed EEG and EDA/HR, respectively, to detect broad emotional states such as positive/negative emotions and valence/arousal without relying on video data. In contrast, Hui~\etal~\cite{hui2018} and Kim~\etal~\cite{kim2020} combined physiological data with video-based methods to detect complex emotions like fear, shame, and guilt. Costa~\etal~\cite{costa2019} and Jiang~\etal~\cite{jiang2019} integrated multi-modal data such as EEG, GSR, and motion data with stimuli like pictures, videos, and music to enhance emotion recognition. While these studies focus primarily on adults and young adults, Badawi~\etal~\cite{badawi2024} addressed older adults by detecting agitation and aggression using physiological signals during indoor activities. 
Usually, detections are made based on biometric signals such as electrodermal activity (EDA), heart rate variability (HRV), blood volume pulse (BVP), and skin temperature (ST). However, most existing works integrate facial emotion recognition~\cite{din2024, badawi2024,saganowski2022} alongside physiological data for improved accuracy or focus on detecting a limited set of emotions rather than a comprehensive range of affective states~\cite{schmidt2018}. Previous studies often depend on deep learning models and cloud-based processing, which significantly limit the feasibility for real-time edge deployment~\cite{wijasena2021, ba2023}. This is of concern, particularly for older adults who may be uncomfortable or unable to participate in such assessments. Thus, the practical use of such a framework in privacy-sensitive settings such as nursing homes or veterans' care facilities is often underexplored~\cite{rhodus2024}.

To address these challenges, we propose an edge-based, non-intrusive emotion detection framework that relies solely on wearable physiological sensor data without requiring cameras or facial analysis. Our proposed work focuses specifically on older adults. These groups are often overlooked in existing studies. Unlike previous research that often integrates facial recognition or video-based methods, our framework uses only physiological signals. Thus, our method eliminates privacy concerns and enables real-time edge-based deployment. This design enhances the practicality and scalability of our method for real-world use in settings such as nursing homes, clinical care, and home environments. Our contribution to this paper is as follows:

\begin{itemize}

    \item We used a data set collected from 40 older adults, where emotions were recorded through physiological signals from the Empatica E4 and Shimmer3 GSR+ wristband.

    \item Participants' facial expression was captured with live video and labeled using iMotion's Facial Expression Analysis (FEA).

    \item We detected the intensity for three distinct emotions: Neutral, Positive, and Negative with only the labeled sensor data. We used machine learning for regression tasks and reported the performance on the test set.

    \item By focusing on classical models rather than deep learning, we ensured computational efficiency, interpretability, and compatibility with edge-based deployment.
    
    \item We demonstrated that emotion recognition can be accurately achieved using only sensor data, eliminating the need for visual or behavioral inputs.
\end{itemize}

The paper is organized as follows: Section \ref{sec_prot} explains the stimulus protocol for emotion and the data recording. The suggested method for emotion detection is presented in Section \ref{proposed_method}. The results of the study and their analysis are presented in section \ref{sec_results}. Section \ref{sec_conc} presents the conclusion with future direction.

\section{Stimulus Protocol TSST and Data Recording}
\label{sec_prot}

In this study, we selected 40 healthy elderly individuals aged between 60 and 80, comprising 28 females and 12 males. Due to the corruption of one participant's data, 39 participants' data were ultimately utilized. Before enrollment, participants underwent screening for any pre-existing medical conditions to ensure an unbiased dataset. The stimulus for emotions was imposed based on \textit{Trier Social Stress Test} (TSST) protocol. 

\begin{figure}[htbp]
    \centering
    \includegraphics[width=\columnwidth]{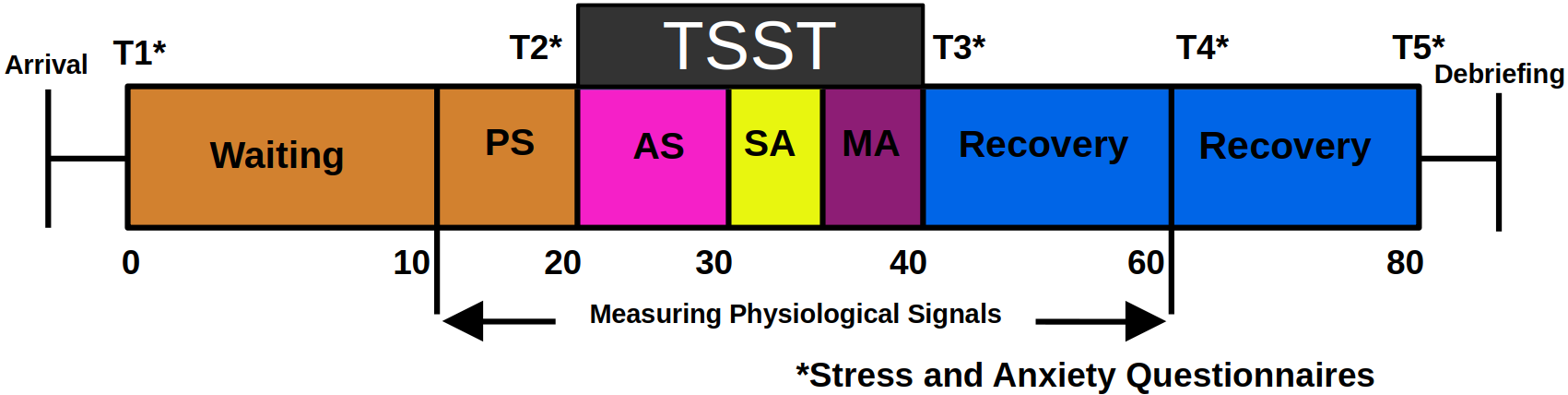}
    \caption{Stimulus Protocol TSST for Capturing Facial Emotion}
    \label{fig:exp_protocol}
\end{figure}

The TSST is a renowned experimental procedure recognized for its effectiveness in causing stress within a naturalistic setting~\cite{Birkett2011}. Thus, we chose the TSST protocol as our external stimulus to impose emotions. Figure~\ref{fig:exp_protocol} presents the stages involved in the experimental process. The TSST protocol encompasses the waiting period, pre-stress phase, stress phase, and recovery phase. During the waiting period, participants complete demographic questionnaires and can ask any questions regarding the study procedures. Baseline data are collected in the pre-stress phase, which directly follows the waiting period. Together, these phases last 20 minutes (T1-T2). The 20-minute stress phase proceeds after the pre-stress phase (T2-T3) and consists of a 10-minute anticipatory stress segment (AS) and a 5-minute segment comprising both speech and mental arithmetic tasks (M). During the AS phase, participants are tasked with delivering a continuous 5-minute speech in front of an audience. The mental arithmetic component requires participants to solve basic addition and subtraction problems under observation. The complexity of these arithmetic tasks increases as participants correctly solve each problem. Finally, the study concludes with two 20-minute recovery phases (T3-T5).

For recording the physiological sensor data, two commercially available smart devices called the Empatica E4 and Shimmer3 GSR+ were used. These devices have EDA, PPG, and ST sensors inbuilt. The recording mode keeps collecting data that can be synchronized after each use. Features collected from the sensors are Yaw, Pitch and Roll from the accelerometer, Temperature, Internal ADC voltage, GSR Resistance, Heart Rate and GSR Conductance.


\section{Proposed Method for Emotion Detection}
\label{proposed_method}

\begin{figure*}
    \centering
    \includegraphics[width=2\columnwidth]{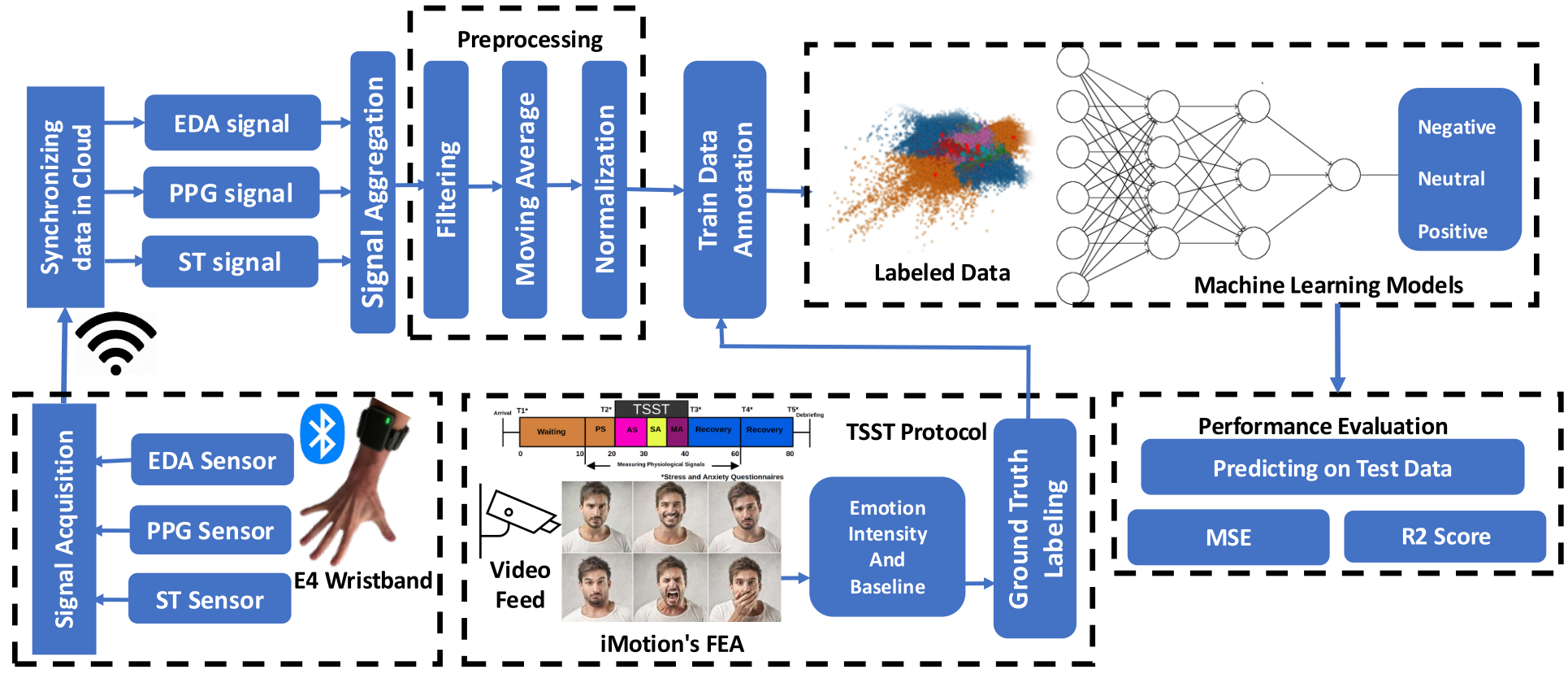}
    \caption{Proposed Emotion Detection Method}
    \label{fig:method}
\end{figure*}

In this section, we will discuss our proposed method for emotion detection with machine learning-based regression models. An overview of the proposed method is shown in Fig.~\ref{fig:method}. The three main components are (i) Data Preprocessing, (ii) Emotion ground truth labeling with iMotion's Facial Expression Analysis (FEA), and (iii) regression with classical machine learning models. In this section, we will explain them in detail.

\subsection{Preprocessing and Ground Truth Labeling}
\label{subsec_gt}
The dataset is preprocessed for labeling after data collection. A 60-sample moving average window was used to scan the time series data during preprocessing. This assisted in eliminating unwanted artifacts and noise. To prevent data leakage, the train and test sets of participants are later divided and normalized independently using z-score normalization. In our study, ground truth labeling for emotion recognition is established using iMotion's Facial Expression Analysis (FEA) module~\cite{imotions_fea_2024}, which employs an advanced implementation of the Facial Action Coding System (FACS). This is a widely used framework in affective computing that can identify facial muscle movements known as Action Units (AUs). Each combination of such action units can correspond to specific emotional expressions. The FEA module leverages computer vision and deep learning algorithms to analyze these AUs in real time. 

\begin{table}[h]
    \centering
    \caption{Emotion Baseline and Percentage Outside 1st Standard Deviation for Labeled Dataset}
    \renewcommand{\arraystretch}{1}
    \setlength{\tabcolsep}{2pt}
    \resizebox{\columnwidth}{!}{%
    \begin{tabular}{lccc|lcc}
        \hline
        \bf Emotion & \bf Baseline & \makecell[c]{\bf \% outside\\ \bf 1st std} & & \bf Emotion & \bf Baseline & \makecell[c]{\bf \% outside\\ \bf 1st std}\\
        \hline
        Joy & -0.6629763 & 24.28 & & Sadness & -0.515366 & 27.38 \\
        Anger & 0.3638698 & 41.11 & & \bf Neutral & -0.0408471 & 34.96 \\
        Surprise & -1.859378 & 7.61 & & \bf Positive & -0.6629763 & 24.28 \\
        Fear & -0.8419589 & 15.12 & & \bf Negative & 0.3638698 & 18.49 \\
        Contempt & -0.1459146 & 33.83 & & Confusion & 0.809446 & 46.93 \\
        Disgust & -0.3184329 & 21.58 & & Frustration & 0.5876518 & 42.75 \\
        \hline
    \end{tabular}
    }
    \footnotesize{\\$\star$ In This work, we will detect only \textit{Neutral, Positive} and \textit{Negative} emotion}
    \label{tab:dataset}
\end{table} \vspace{-0.15in}

\begin{figure}[htbp]
    \centering
    \includegraphics[width=\columnwidth]{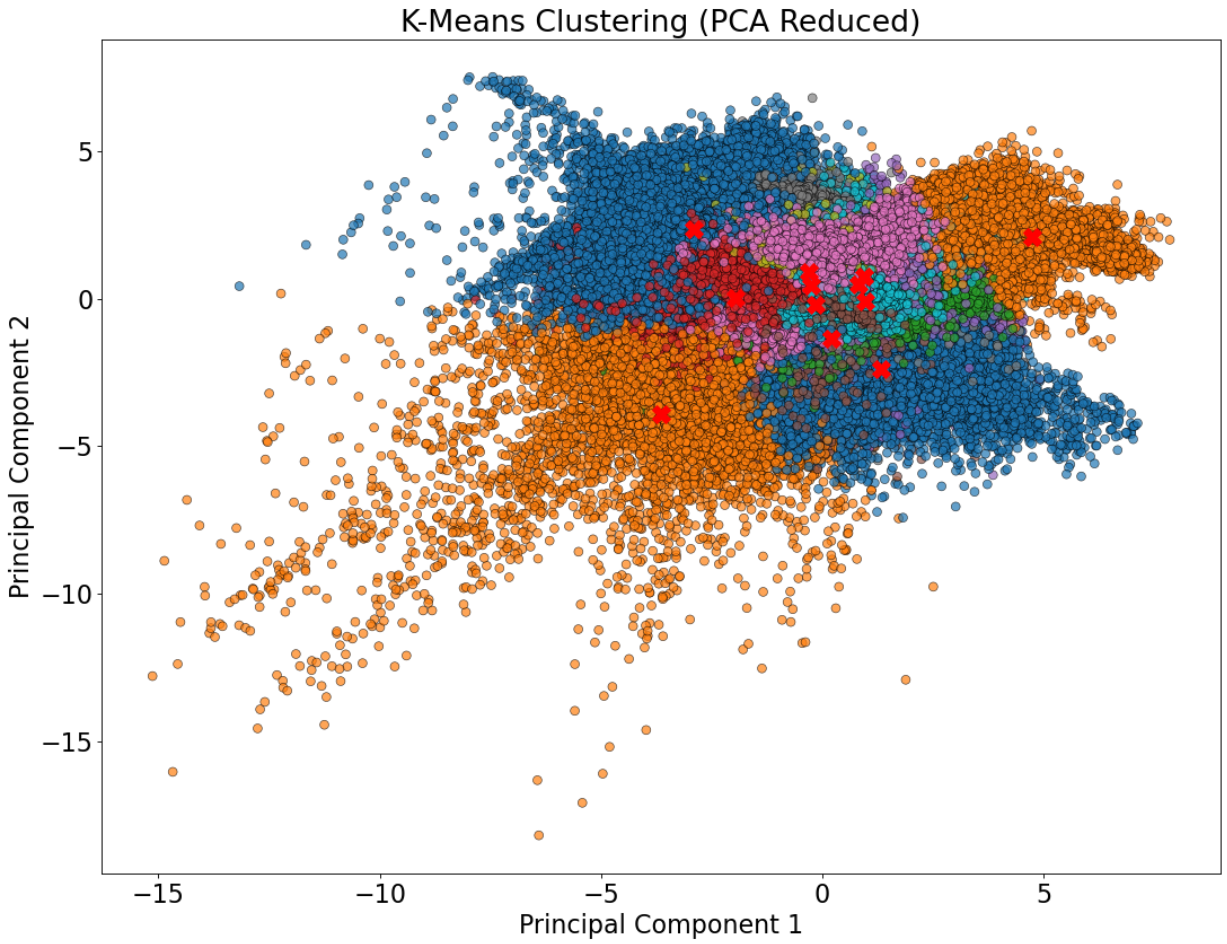}
    \caption{2D Projection of the Labeled Dataset with PCA Showing the Cluster Centers where Each Color Represents Different Emotion}
    \label{fig:dataset}
\end{figure}

This automated analysis ensures a standardized, objective method for labeling emotions, reducing potential bias associated with self-reporting. By synchronizing physiological sensor data (EDA, BVP, and IBI) with facial expression-derived emotion labels, we created a robust dataset that enables training and validation of our edge-based emotion recognition framework. After the labeling, a distribution of Emotion intensity is shown in Table~\ref{tab:dataset}. Again, Fig.~\ref{fig:dataset} shows a 2D projection of all the emotion classes after labeling. The centroid of the cluster shows how all the emotional states can overlap and are hard to distinguish. In this work, we plan to detect only 3 sets of emotions: \textit{Positive, Negative} and \textit{Neutral}.

\subsection{Machine Learning Models for Regression}
\label{ml}
\begin{figure}[htbp]
    \centering
    \includegraphics[width=1.05\columnwidth]{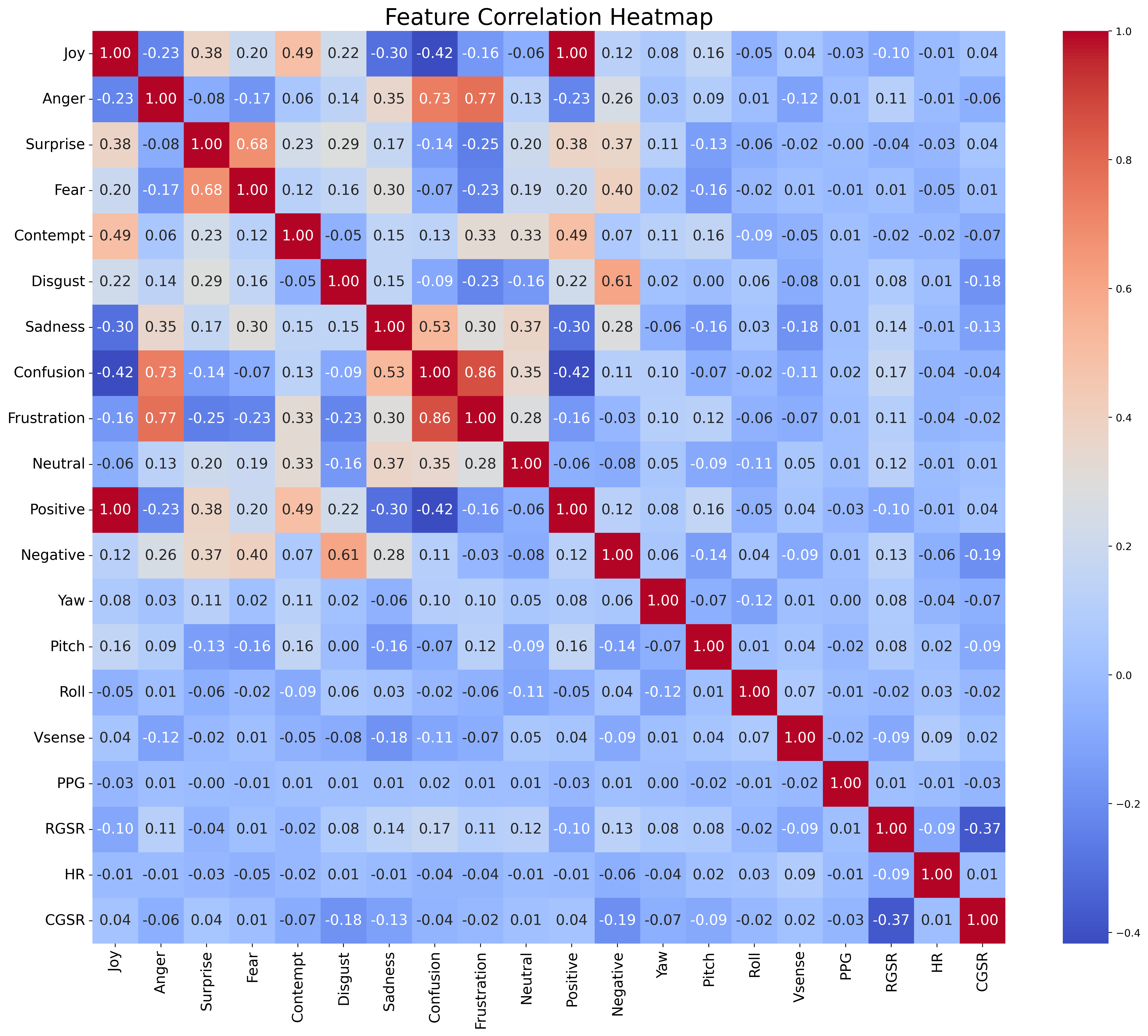}
    \caption{Correlation Heatmap of the Features and Emotion of Dataset}
    \label{fig:heatmap}
\end{figure}

To evaluate the performance of our framework, we implemented seven classical machine learning methods with one 10-layer Multi-Layer Perceptron (MLP) and a 3-hidden-layer Dense Neural Network. The correlation heatmap shown in Fig.~\ref{fig:heatmap} summarizes that traditional statistical modeling will make it difficult to make an emotion detection method. Thus, machine learning models that can handle such small correlations will prove to be well-suited for such problems. 

The most simplistic regression model is the linear regression. It assumes a direct linear relationship between physiological features and emotional intensity. It estimates coefficients that best fit the data by minimizing the sum of squared errors. Ridge Regression is a linear model that incorporates L2 regularization, which penalizes large coefficients to prevent overfitting. Bayesian Ridge Regression extends Ridge Regression by incorporating a probabilistic framework, which assigns prior probability distributions to the model coefficients. 

K-Nearest Neighbors (KNN) Regression is a non-parametric model that estimates emotion intensity based on the average intensity values of the K most similar physiological responses. This approach assumes that individuals with similar physiological reactions experience similar emotional intensities. Decision Tree Regression is a rule-based model that recursively splits the data into subsets based on the feature values that minimize the prediction error at each node. Another prominent method is the Random Forest Regression which is an ensemble learning method. It constructs multiple decision trees during training and averages their predictions for a more accurate and robust outcome.

Gradient Boosting Regression is a more complex model. It is an advanced ensemble technique that sequentially builds multiple weak learners (typically decision trees), where each new tree corrects the errors of the previous ones. This method reduces both bias and variance, leading to highly accurate predictions. Multi-Layer Perceptron (MLP) Regression is a type of feedforward neural network consisting of multiple hidden layers. Each layer processes the input data using weighted connections and non-linear activation functions. Finally, A Dense Neural Network (DNN) consists of multiple fully connected layers that transform input features through non-linear activation functions. By learning hierarchical representations, DNNs can capture complex interactions between physiological signals and emotional intensity levels.

\subsection{Model Training and testing}
\label{subsec_train}
As we had participant-specific data, there were multiple ways to generalize the machine learning models based on the dataset. To avoid data leakage, we split the dataset based on participants before normalizing them. We performed \textit{Leave One Sample Out (LOSO)} cross-validation keeping one subject for testing and others for training. We used some widely used regression models as our base machine learning model mentioned in Section~\ref{ml}. For each of the regression models, the set of hyperparameters used is shown in Table~\ref{tab:hyp}. The values for these parameters were chosen based on the best-performing set on test data after multiple trial and error.

\begin{table}[h]
    \centering
    \caption{Hyper-parameters Used to Train Machine Learning Models}
    \renewcommand{\arraystretch}{1}
    \setlength{\tabcolsep}{8pt}
    \resizebox{\columnwidth}{!}{%
    \begin{tabular}{ll}
        \toprule
        \bf Machine Learning Model & \bf Hyper-parameters\\
        
        \midrule
        Random Forest & \makecell[l]{no of estimators=100,\\criterion=squared error}\\
        
        \midrule
        Dense Neural Network & \makecell[l]{no of hidden layers=3, loss=mse,\\output layer activation=ReLU,\\optimizer=Adam, epochs=50,\\Learning rate = 0.0001}\\
        
        \midrule
        Decision Tree & \makecell[l]{criterion=squared error}\\

        \midrule
        K-Nearest Neighbors (KNN) & \makecell[l]{no of neighbors=3, weights=uniform,\\distance metrics=minkowski}\\

        \midrule
        Gradient Boosting & \makecell[l]{no of estimators=100, Learning rate=0.1,\\loss=squared error,criterion=friedman mse}\\

        \midrule
        Multi Layer Perceptron & \makecell[l]{no of hidden layers=10, loss=mse,\\output layer activation=ReLU,\\optimizer=Adam, Learning rate = 0.001,\\ maximum iteration=500}\\

        \midrule
        Ridge Regression & \makecell[l]{maximum iteration=1000,\\tolerance=$10^{-4}$, solver=svd}\\

        \midrule
        Bayesian Ridge Regression & \makecell[l]{maximum iteration=1000,\\tolerance=$10^{-3}$, $\alpha_1, \alpha_2, \lambda_1, \lambda_2 = 10^{-6}$}\\

        \midrule
        Linear Regression & \makecell[l]{N/A}\\
        \bottomrule
    \end{tabular}
    \label{tab:hyp}
    }
\end{table}


\section{Result Analysis and Discussion}
\label{sec_results}
In this section, we will evaluate the performance of machine learning models with our dataset with respect to the estimated ground truth mentioned in Section~\ref{subsec_gt}. We have evaluated the model’s performance using $r^2$ score and MSE with the equations shown in Eq.~\eqref{eq:r2} and~\eqref{eq:mse} where, $y_i$ = actual value, $\hat{y}_i$= predicted value and $\bar{y}_i$ = mean of actual value. The $r^2$ score of a model says how much of the variability in the target that model can explain in terms or normalized distance. Again, MSE refers to the squared difference between the model's predicted values and the actual values. The scores are expressed as the mean of all folds for LOSO cross-validation.

\begin{equation}
    R^2\ score = 1 - \frac{\sum_{i=1}^{n} (y_i - \hat{y}_i)^2}{\sum_{i=1}^{n} (y_i - \bar{y})^2}
\label{eq:r2}
\end{equation}

\begin{equation}
    \text{MSE} = \frac{1}{n} \sum_{i=1}^{n} (y_i - \hat{y}_i)^2
\label{eq:mse}
\end{equation}

The results indicate that Random Forest outperforms other models as evidenced by its highest $r^2$ scores in Table~\ref{tab:r2_scores} and lowest Mean Squared Error (MSE) in Table~\ref{tab:mse_comparison} across all emotion categories. That is because the ensemble nature of Random Forest helps reduce overfitting and enhances generalization. This indicates that the Random Forest model was able to explain a substantial portion of the variance in emotional states with reasonable confidence.




\begin{table}[h]
    \centering
    \caption{$r^2$ Scores for Various ML Models (Higher is better)}    \renewcommand{\arraystretch}{1}
    \setlength{\tabcolsep}{8pt}
    \resizebox{\columnwidth}{!}{%
    \begin{tabular}{lccc}
        \toprule
        \bf ML Model & \makecell[c]{\bf Negative} & \makecell[c]{\bf Neutral} & \makecell[c]{\bf Positive} \\
        \midrule
        Random Forest & 0.782 & 0.7636 & 0.8033 \\
        Dense Network & 0.5657 & 0.5339 & 0.5813 \\
        Decision Tree & 0.5377 & 0.5253 & 0.5762 \\
        K-Nearest Neighbors (KNN) & 0.5217 & 0.5021 & 0.5673 \\
        Gradient Boosting & 0.3545 & 0.3644 & 0.2997 \\
        Multi Layer Perceptron & 0.3088 & 0.2568 & 0.2655 \\
        Ridge Regression & 0.0752 & 0.0371 & 0.0527 \\
        Bayesian Ridge Regression & 0.0752 & 0.0370 & 0.0526 \\
        Linear Regression & 0.0752 & 0.0370 & 0.0526 \\
        \bottomrule
    \end{tabular}

    \label{tab:r2_scores}
    }
\end{table}

In contrast, simpler models such as Ridge Regression, Bayesian Ridge Regression, and Linear Regression perform poorly. They showed significantly lower $r^2$ scores with high MSE. This implies that linear models fail to capture the intricate dependencies between physiological signals and emotion states. Since physiological signals are often highly nonlinear and affected by multiple interacting factors, simpler models struggle to provide meaningful predictions.


\begin{table}[h]
    \centering
    \caption{MSE for Various ML Models (Lower is better)}
    \renewcommand{\arraystretch}{1}
    \setlength{\tabcolsep}{8pt}
    \resizebox{\columnwidth}{!}{%
    \begin{tabular}{lccc}
        \toprule
        \bf ML Model & \makecell[c]{\bf Negative} & \makecell[c]{\bf Neutral} & \makecell[c]{\bf Positive} \\
        \midrule
        Random Forest & 0.0006 & 0.0014 & 0.0019 \\
        Dense Network & 0.0012 & 0.0029 & 0.0030 \\
        Decision Tree & 0.0012 & 0.0030 & 0.0030 \\
        K-Nearest Neighbors (KNN) & 0.0013 & 0.0031 & 0.0044 \\
        Gradient Boosting & 0.0017 & 0.0042 & 0.0050 \\
        Multi Layer Perceptron & 0.0018 & 0.0047 & 0.0052 \\
        Ridge Regression & 0.0025 & 0.0060 & 0.0067 \\
        Bayesian Ridge Regression & 0.0025 & 0.0060 & 0.0067 \\
        Linear Regression & 0.0025 & 0.0060 & 0.0067 \\
        \bottomrule
    \end{tabular}
    \label{tab:mse_comparison}
    }
\end{table}

Neural network-based models, such as the Dense Network and Multi-Layer Perceptron (MLP), perform moderately well. While they achieve better $r^2$ scores than linear models, their higher MSE values suggest a higher degree of variance in their predictions. However, their ability to outperform purely linear models suggests that with further optimization and larger datasets, deep learning approaches could become more competitive in emotion detection from physiological signals with more complexity as that would make them harder to deploy in the edge.


\section{Conclusion}
\label{sec_conc}
This study demonstrates the feasibility of emotion detection using physiological signals alone that eliminates the need for intrusive facial recognition methods. Our results show that classical machine learning models can perform better in predicting emotion intensities. These findings validate the effectiveness of wearable sensor data in capturing emotional states that can offer a privacy-preserving and edge-compatible solution for real-world applications. This has significant implications for individuals with Alzheimer’s Disease and Related Dementia (ADRD), veterans with Post-Traumatic Stress Disorder (PTSD), and other cognitive impairments. This approach can be integrated into hospital and assisted living settings to monitor emotional well-being without requiring active participation from individuals. This is especially beneficial for patients with limited verbal or cognitive abilities. Future work can expand on this foundation by incorporating deep learning models, transfer learning, and larger datasets to enhance performance further. This study lays the groundwork for real-time, privacy-conscious emotion detection systems which is viable both for healthcare monitoring and human-centered AI.



\bibliographystyle{ACM-Reference-Format}
\bibliography{references}

\end{document}